\documentclass[twocolumn,twocolappendix]{aastex63}
\usepackage{amsmath}
\usepackage{amssymb}
\usepackage{bm}
\usepackage{graphicx}
\usepackage{color}
\usepackage{float}
\usepackage{tabularx}
\usepackage{ragged2e}
\usepackage{graphicx}
\usepackage{subfigure}
\usepackage{booktabs}
\usepackage{multirow}

\definecolor{darkgreen}{rgb}{0,0.35,0}
\definecolor{blue}{rgb}{0,0,1}

\usepackage{CJKutf8}            

\newcommand{\be}{\begin{eqnarray}}
\newcommand{\ee}{\end{eqnarray}}

\newcommand\as{\bgroup\markoverwith{\textcolor[rgb]{.5, 0, .6}{\rule[0.5ex]{8pt}{1.5pt}}}\ULon}

\defcitealias{Li2021}{L21}

\shorttitle{Migration of Accreting Planets}
\shortauthors{Pan et al.}

\begin{document}

\title{Concurrent Accretion and Migration of Giant Planets in their Natal Disks with Consistent Accretion Torque (II): Parameter Survey and Condition for Outward Migration}

\author[0009-0002-7003-8883]{Junpeng Pan\begin{CJK*}{UTF8}{gbsn}(潘俊鹏)\end{CJK*}}
\affiliation{Shanghai Astronomical Observatory, Chinese Academy of Sciences, Shanghai 200030, China}
\affiliation{University of Chinese Academy of Sciences, Beijing 100049, China}
\author[0000-0002-7329-9344]{Ya-Ping Li\begin{CJK*}{UTF8}{gbsn}(李亚平)\end{CJK*}}
\affiliation{Shanghai Astronomical Observatory, Chinese Academy of Sciences, Shanghai 200030, China}
\author[0000-0003-3792-2888]{Yi-Xian Chen\begin{CJK*}{UTF8}{gbsn}(陈逸贤)\end{CJK*}}
\affiliation{Department of Astrophysics, Princeton University, Princeton, NJ 08544, USA}
\author[0000-0001-9564-6186]{Shigeru Ida\begin{CJK*}{UTF8}{gbsn}(井田茂)\end{CJK*}}
\affiliation{Earth-Life Science Institute, Institute of Science Tokyo, Tokyo 152-8550, Japan}
\affiliation{Department of Astronomy, School of Science, Westlake University, Hangzhou, Zhejiang 310030, China}
\author[0000-0001-5466-4628]{Douglas N. C., Lin \begin{CJK*}{UTF8}{gbsn}(林潮)\end{CJK*}}
\affiliation{Department of Astronomy \& Astrophysics, University of California, Santa Cruz, CA 95064, USA}
\affiliation{Department of Astronomy, School of Science, Westlake University, Hangzhou, Zhejiang 310030, China}
\affiliation{Institute for Advanced Studies, Tsinghua University, Beijing, 100084, China}


\correspondingauthor{Ya-Ping Li}
\email{liyp@shao.ac.cn}

\begin{abstract}

Migration typically occurs during the formation of planets and is closely linked to the planetary formation process. 
In classical theories of non-accreting planetary migration, both type I and type II migration typically result in inward migration, which is hard to align with the architecture of the planetary systems.
In this work, we conduct systematic, high-resolution 3D/2D numerical hydrodynamic simulations to investigate the migration of an accreting planet. 
Under different disk conditions, we compared the dynamical evolution of planets with different planet-to-star mass ratios. 
We find that accretion of planets can significantly diminish the inward migration tendency of planets, or even change the migration direction. 
The migration of low-/high-mass planets is classified as Type I/II inward migration, respectively, while intermediate-mass planets, which have the strongest accretion, show an outward migration trend. 
We confirm that the outward migration is mainly attributed to the positive torque from the azimuthal asymmetric structures around the accreting planet, similar to \citet{Li2024}.  
The termination of planetary mass growth is thus synonymous with the transition from outward to inward migration.
For the high viscosity $\alpha=0.04$ and disk aspect ratio height $h_0=0.05$ cases, the mass ratio range for planetary outward migration is $1\times10^{-4}\lesssim q\lesssim4\times10^{-3}$. 
For the low viscosity case with $\alpha=0.001$, and/or the low disk aspect ratio cases $h_0=0.03$, the mass ratio range for the outward migration will shift toward the lower end. Our parameter survey reveals that a simple gap opening parameter determines the outward migration condition; details of the analytical interpretation are presented in \citet{Ida2025}.

\end{abstract}

\section{Introduction}\label{sec:intro}
Up to $\sim6000$ exoplanets have been detected up to now, and they are categorized into three major classes: super-Earth, hot Jupiter and cold Jupiter \citep{ZhuDong2021}. The close-in orbits of hot Jupiter \citep{Lin1996} and super-Earths \citep{Borucki2011} prompted the hypothesis of planetary migration. Planets embedded in protoplanetary disks exchange angular momentum with the disk material, leading to changes in their orbits, as process know as migration. A similar mechanism is expected to operate for stellar-mass black holes (sBHs) embedded in active galactic nucleus disks \citep{Tagawa2020, WuChenLin2024}. 
\par
In the widely accepted theory, migration is classified into Type I and Type II migration. For low-mass planets, their perturbation on disk is weak, thus correspond to the Type I migration. The planet experiences torques from the density waves excited by itself. The outer spiral arm of the wave provides a negative torque, while the inner spiral arm provides a positive torque. Generally, the asymmetry between the outer and inner spiral arms results in a net negative torque, driving the planet to migrate inward \citep{PapaloizouLin1984, LinPapaloizou1993, Kley_Nelson_2012, Paardekooper2023}. 
For giant planets, classical theory suggests that the perturbation on the disk is substantial enough to open a clear gap near its orbit, 
thus transitioning the migration to Type II, where the planet migrates inward at a velocity $v_{\rm r}=-3\nu/2r$ that is comparable to the viscous radial velocity of the background disk \citep{LinPapaloizou1986b,Bryden2000}.
However, recent simulations indicate that the gap is not completely devoid of gas, as residual flow across it sustains a steady density profile \citep[e.g.,][]{Duffell2014,Fung2014,DurmannKley2015,Robert2018,Chen2020b,Li2023}. Consequently, the  migration rate of moderate mass gap-opening planets aligns more closely with a more generalized migration prescription \citep{Kanagawa2018}. 
The migration rate from these simulations still poses a challenge to the retention of cold Jupiters when compared with 
the observed period distribution \citep{Nelson2000,Ida2013}. 
Some scenarios have been investigated to identify processes which may lead to 
the slowing down of migration speed or even the reversal of migration direction \citep[e.g.,][]{Masset2006,Lega2014,Chen2020b,Dempsey2021,Liu2025}. 
Nevertheless, these theories of migration focus only on non-accreting planets, in which case migration rates are inward. 

Some studies have used numerical hydrodynamical simulations to investigate the accretion dynamics of giant planets or companions \citep{DAngelo2003, DAngelo2008, Machida2010, Bodenheimer2013, Li2021, Li2023, Choksi2023}, but they rarely discuss the impact on migration. 
Some research has also conducted analytical or N-body simulations to study the migration dynamics of accreting planets \citep{TanigawaTanaka2016,Tanaka2020,Wang2020,Jiang2023,Robert2018}. 
Only recently have detailed hydrodynamical simulations emerged that self-consistently model accretion and migration torques \citep{Li2024,Laune2024,Wu2024}.


\cite{Li2024} carried out a few 3D and 2D global hydrodynamical simulations to study the concurrent migration and accretion of an accreting Jupiter-mass planet in a range of viscosity parameters.  
It was found that accreting planets tend to migrate outward when the disk viscosity is high enough.
\cite{Laune2024} performed a series of 2D global hydrodynamical simulations with an iterative method that enables the disk to reach a quasi-steady state in a relatively short period of time. They consistently found that accreting planets exhibit an outward migration trend in a similar parameter regime. However, the parameter space explored to date remains limited, hindering a comprehensive understanding of migration and accretion processes. 

Crucially, accretion's influence on migration is driven by gravitational torques from asymmetric structures in circumplanetary disks (CPDs; \citealt{Li2024}), which exhibit strong sensitivity to both disk and planetary parameters \citep[e.g.,][]{Li2023}. To address this dependence, we ystematically explore migration and accretion processes for an accreting planet across a wide parameter space -- varying planet masses, disk viscosities, and disk aspect ratios in a self-consistent way in this work. This extensive parameter study enables the development of a comprehensive analytical framework for accreting-planet migration dynamics \citep{Ida2025}, directly applicable to future population synthesis studies.


In this work, we carry out 3D high-resolution, global hydrodynamical simulations to study the evolution of accreting planet's (or sBH's) migration and accretion in a large-parameter regime in a self-consistent way. 
The rest part of the paper is organized as follows. We describe the method of simulation and data analysis in Section \ref{sec:method}. The primary results are presented in Section \ref{sec:result}. Section \ref{sec:summary} shows the summary and discussions.

\begin{table*}
\centering
\caption{Model Parameters and Simulation Results}
\label{tab:para}
\begin{tabularx}{\textwidth}{c c c c c c c c c c c c c} 
\hline
\hline
\textbf{Models} &
$q$ &
\textbf{$\dot{m}_{\rm p}$} & 
\textbf{$\dot{a}_{\rm p,grav} /a_{\rm p}$} & 
\textbf{$\dot{a}_{\rm p,acc} /a_{\rm p}$} & 
\textbf{$\Gamma_{\rm grav}$}&
\textbf{$\Gamma_{\rm acc}$}&
\textbf{$\Gamma_{\rm tot}$}&
\textbf{$\dot{a}_{\rm p}/a_{\rm p}$} & 
Remarks &
\\ 
&&
\textbf{$(\Sigma_0 r_0^2 \Omega_0)$} & 
\textbf{$(\dot{m}_{\rm d}/M_{*})$} & 
\textbf{$(\dot{m}_{\rm d}/M_{*})$} & 
\textbf{$(\Sigma_0r_0^4\Omega_0^2)$}&
\textbf{$(\Sigma_0r_0^4\Omega_0^2)$}&
\textbf{$(\Sigma_0r_0^4\Omega_0^2)$}&
\textbf{$(\dot{m}_{\rm d}/M_{*})$} & 
\\ \hline
mp5e5&$5\times10^{-5}$&$4\times10^{-4}$&-39&6&$-1\times10^{-6}$&$1.5\times10^{-7}$&$8.5\times10^{-7}$&-33&
\multirow{8}{2cm}{\centering $h_0=0.05$ \\ $\alpha=0.04$ \\ $3\rm{D}$}\\ 
mp1e4&$1\times10^{-4}$&$4.3\times10^{-4}$&32&-2&$1.4\times10^{-6}$&$-8.8\times10^{-8}$&$1.3\times10^{-6}$&30\\ 
mp2e4&$2\times10^{-4}$&$5.8\times10^{-4}$&29&$\sim 0$&$5\times10^{-6}$&$\sim 0$&$5\times10^{-6}$&30\\
mp5e4&$5\times10^{-4}$&$7\times10^{-4}$&44&-4&$1.4\times10^{-5}$&$-1.3\times10^{-6}$&$1.3\times10^{-5}$&40\\
mp1e3&$1\times10^{-3}$&$8.9\times10^{-4}$&57&-10&$2.4\times10^{-5}$&$4.2\times10^{-6}$&$2.8\times10^{-5}$&50\\
mp2e3&$2\times10^{-3}$&$8.9\times10^{-4}$&31&1&$3\times10^{-5}$&$1\times10^{-6}$&$3.1\times10^{-5}$&32\\
mp4e3&$4\times10^{-3}$&$9\times10^{-4}$&-1&7&$-3\times10^{-6}$&$2.1\times10^{-5}$&$1.8\times10^{-5}$&6\\
mp8e3&$8\times10^{-3}$&$7.6\times10^{-4}$&-40&11&$-1.6\times10^{-4}$&$4.4\times10^{-5}$&$-1.2\times10^{-4}$&-29\\
\hline
mp3e5a&$3\times10^{-5}$&$2\times10^{-5}$&900&-400&$3.2\times10^{-7}$&$-1.4\times10^{-7}$&$1.8\times10^{-7}$&530&
\multirow{5}{2cm}{\centering $h_0=0.05$ \\ $\alpha=0.001$ \\ $2\rm{D}$}\\
mp1e4a&$1\times10^{-4}$&$2\times10^{-5}$&220&-10&$2.6\times10^{-7}$&$1.2\times10^{-8}$&$2.5\times10^{-7}$&210&\\
mp3e4a&$3\times10^{-4}$&$2\times10^{-5}$&-15&35&$-5\times10^{-8}$&$1.2\times10^{-7}$&$8\times10^{-8}$&20&\\
mp1e3a&$1\times10^{-3}$&$2\times10^{-5}$&-120&-10&$-1.3\times10^{-6}$&$-1\times10^{-7}$&$-1.4\times10^{-6}$&-120&\\
mp2e3a&$2\times10^{-3}$&$6\times10^{-6}$&-80&20&$-1.9\times10^{-6}$&$5\times10^{-7}$&$-1.4\times10^{-6}$&-60&\\
mp1e3a*&$1\times10^{-3}$&$2\times10^{-5}$&-200&-30&$2.4\times10^{-6}$&$3.6\times10^{-7}$&$2.8\times10^{-6}$&-230&3D\\
\hline
mp3e4h&$3\times10^{-4}$&$2.8\times10^{-4}$&100&60&$5\times10^{-6}$&$3\times10^{-6}$&$8\times10^{-6}$&160&
\multirow{3}{2cm}{\centering $h_0=0.03$ \\ $\alpha=0.04$ \\ $2\rm{D}$}\\
mp1e3h&$1\times10^{-3}$&$2.8\times10^{-4}$&60&20&$1\times10^{-5}$&$3.3\times10^{-6}$&$1.3\times10^{-5}$&80&\\
mp2e3h&$2\times10^{-3}$&$2.9\times10^{-4}$&15&10&$4\times10^{-6}$&$2.7\times10^{-6}$&$6.7\times10^{-6}$&25&\\
\hline
\hline
\end{tabularx}
\vspace{1cm}
\begin{minipage}{\textwidth}
\footnotesize \justifying
\textbf{Note. }For 3D models, the disk viscosity is fixed as $\alpha=0.04$, and the disk scale height is $h_0=0.05$, while the planet-to-star mass ratio $q\equiv m_{\rm p}/M_*$ may vary for different models. 
$\dot{m}_{\rm d}$ is the disk accretion rate supplied from the outer boundary. A positive/negative torque and $\dot{a}_{\rm p}/a_{\rm p}$ means that the planet migrates outward/inward with time. 
\end{minipage}
\end{table*}

\section{Method}\label{sec:method}

Most of the model setup and numerical methods are the same as in \cite{Li2024}. 
Here we only briefly introduce some key ingredients of the models.
Our simulations can be also applied to an accreting stellar-mass BH embedded in AGN disks. In the following, we use the planet to refer to the embedded object in the disk.

In this study, we perform global 3D and 2D hydrodynamical simulations of an accreting planet embedded in a thin, non-self-gravitating protoplanetary disk using \texttt{Athena++} \citep{Stone2020}. 
We numerically solve the continuity equation and the equation of motion in the co-rotating coordinate system of the planet, with the origin of which located at the host star with mass $M_*$.
For 3D models, the simulations are performed using spherical coordinates $(r,\theta,\phi)$, with the initial density distribution defined in cylindrical radius $(R=r\sin\theta)$ and vertical height $(z=r\cos\theta)$, following the setup in \cite{Nelson2013}:
\begin{equation}
    \rho=\rho_0\bigg( \frac{R}{r_0}\bigg)^p{\rm exp}\bigg[\frac{GM_*}{c_{\rm s}^2}\bigg(\frac{1}{\sqrt{R^2+z^2}}-\frac{1}{R}\bigg)\bigg],
\end{equation}

\par
In calculating the gravitational of the planet at $\bm{r}$, we use a smoothed potential of the form 
\begin{equation}
    \Phi_{\rm p}=-\frac{Gm_{\rm p}}{(|\bm{r}_{\rm p}-\bm{r}|^2+\epsilon^2)^{1/2}}+q\Omega_{\rm p}^2\bm{r}_{\rm p}\cdot\bm{r},
\end{equation}
where $\bm{r}_{\rm p}$ indicates the location of the planet, $\epsilon$ is the softening length, and $\epsilon=0.1R_{\rm H}$ is adopted unless otherwise stated, $q=m_{\rm p} /M_*$ is the mass ratio between planet and the host star. 
In this study, we consider a planet in a fixed circular orbit with $r_{\rm p}=a_{\rm p}$ where $a_{\rm p}$ is the orbital semimajor axis of planet. 
The angular velocity of the planet $\Omega_{\rm p}=\sqrt{GM_*/a_{\rm p}^3}$.

In order to avoid some numerical instabilities,  the planet’s gravitational potential is gradually ``turned-on'' over ten orbital period, allowing the disk to respond to the planet potential smoothly. 
For some massive planet cases, such as models \texttt{mp2e3}, \texttt{mp4e3} and \texttt{mp8e3} in Table \ref{tab:para}, the simulations are performed by restarting the simulation with a less massive planet, and the softening radius is adjusted to $0.1\ R_{\rm H}$ accordingly. For some less massive planets (e.g., models \texttt{mp5e4}, \texttt{mp2e4}, \texttt{mp1e4}), the mesh refinement level is increased to ensure the similar resolution within the Hill sphere. This kind of treatment can avoid the numerical issues for massive planet accretion, and save the computation time. 

We adopted the conventional $\alpha$ prescription for the kinematic viscosity \citep{ShakuraSunyaev1973} with $\nu=\alpha H^2\Omega$. 
The disk viscosity of protoplanetary disks (PPDs) is usually on the order of $\alpha\sim10^{-3}$, but it could be much higher for the innermost region close to the host star and for the AGN disks. 
Among all our 3D simulations, we set a high viscosity condition $\alpha=0.04$ and $h_{0}=0.05$, unless otherwise noted.
We also performed a few 2D simulations for $\alpha=0.001$ or $h_{0}=0.03$, in which the setups are the same as the 2D runs in \cite{Li2024}. The parameters explored are summarized in Table \ref{tab:para}.

In all cases, we take the natural units of $G=M_*=r_0=1$. The planet is fixed at a semimajor axis of $a_{\rm p}=r_0=1$ with $\Omega_{\rm p}=\Omega_0=1$.

\subsection{Planetary Accretion}

To account for the active gas accretion of the planet, we adopted the sinkhole method following \cite{Li2024}, which is based on previous studies \citep{Kley2001, DAngelo2003,Li2021,Li2021b}. 
In this approach, gas is removed at a rate $\eta$ (in units of the local Keplerian frequency $\Omega_0$) from cells within a radius $r_{\rm a}$ around the planet. 
We remove a uniform fraction of mass in each cell within $\delta r<r_{\rm a}$ each numerical time step, where $\delta r$ represents the distance to the planet. The accretion rate onto the planet $\dot{m}_{\rm p}$ is computed with
\begin{equation}
    \dot{m}_{\rm p}=\int_{\delta r<r_{\rm a}}\eta \rho dV,
\end{equation}
where $dV$ is the volume of the sink cell. 
In our simulations, we set $r_{\rm a}=\epsilon=0.1R_{\rm H}$ and $\eta=50\Omega_0$. 
According to \cite{Li2024}, a smaller accretion radius does not significantly affect our results. 
Such a removal rate can ensure the convergence of the planetary accretion rate but does not induce significant eccentric instability around the sink cell, which has been further confirmed by the negligible contribution of the gravitational torque from the material around the sink cell. In addition, we have also performed some simulations with smaller $\eta$, which show an insignificant difference for the migration rates of the planet when the accretion rate of the planet converges with $\eta$.

\subsection{Mesh Refinement and Boundary Conditions}

We use 160 uniform radial grids spaced between $r_{\rm min}=0.5r_0$ and $r_{\rm max}=2.5r_0$; and 512 uniform grids in azimuth for the base level. 
In order to save computational expense, only half a disk above the midplane with 16 root grids is used after considering the symmetry.
In all of our simulations, we have fixed the planetary location and the planetary mass. 
For giant planets, especially, the radial computational domain of the disk is not sufficiently far away from the planet location to avoid boundary effects. 
Therefore, the radial grids are adjusted to $r_{\rm min}=0.3$ and $r_{\rm max}=3.5$. 
Four or five levels of static mesh refinement (SMR) are adopted within the region $\delta r<R_{\rm H}$. 
For models with $q\geq1\times10^{-3}$, we adopt 4 levels of SMR. In order to resolve the small mass ratio cases, we adopt 5 levels of SMR when $q<1\times10^{-3}$. For the minimum mass ratio model \texttt{mp5e5}, the Hill radius of the planet is as small as $R_{\rm H}\simeq0.026r_0$. The Hill radius can thus be resolved by 64 grids in each dimension. For comparison, the Hill radius of the model \texttt{mp1e3} ($q=10^{-3}$) can be covered by 90 grids in each dimension. Therefore, this adjustment in resolution is sufficient to capture the gas dynamics closest to the planet while also saving computational resources.


The boundary condition with wave-killing at the inner and outer radial edges is implemented following \citet{Dempseyetal2020,Li2024}. This kind of boundary can ensure the disk accretion rate outside the planetary orbit to match the desired mass flux $\dot{m}_{\rm d}$ at the outer edge, and the disk materials inside the planetary orbit accrete onto the host star with a jump self-consistently controlled by the planetary accretion rate. 


\begin{figure*}
    \centering
    \includegraphics[width=\linewidth]{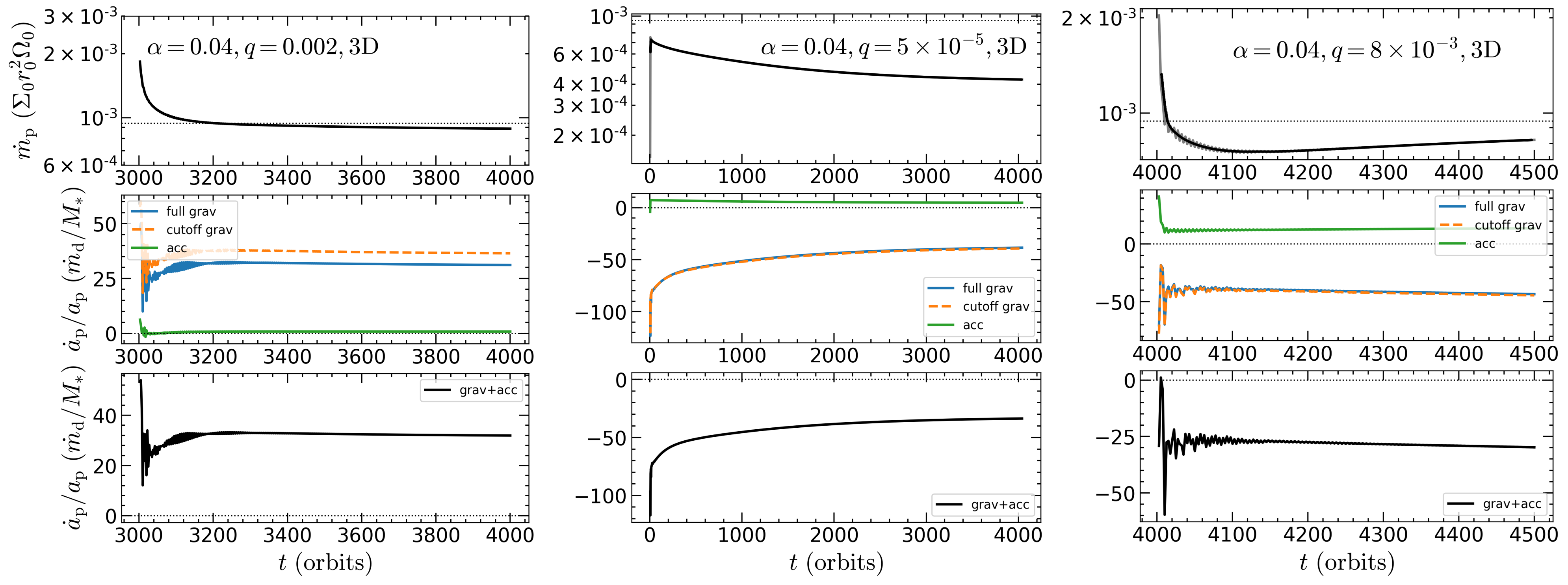}
    \caption{
    The first row shows the evolution of planetary accretion rate $\dot{m}_{\rm p}$.
    The middle and third row show the semi-major axis evolution $\dot{a}_{\rm p}/a_{\rm p}$ due to the gravitational and accretion torque and their sum. 
    Different columns correspond to models with different mass ratio $q$.
    The dashed line in first row shows the disk accretion rate $\dot{m}_{\rm d}$ from outer boundary. 
    The blue and green line in middle row show the migration rate $\dot{a}_{\rm p}/a_{\rm p}$ contributed by gravitational and accretion torque, respectively. 
    The orange dashed line represents the gravitational torque in which the sinkhole part $(\delta r<r_{\rm a})$ has been excluded.
    }
    \label{fig:1}
\end{figure*}

\subsection{Migration Rate}
The gravitational interaction between the disk and the planet drives the planetary migration. The gravitational force on the planet is 
\begin{equation}
    \bm F_{\rm grav} = \int\rho\nabla\Phi_{\rm p}dV,
\end{equation}
where $\Phi_{\rm p}$ is the gravitational potential of the planet. 
The torque resulting from gravitational forces is $\bm \Gamma_{\rm grav}=\bm r_{\rm p}\times\bm F_{\rm grav}$ and thus the angular momentum changes by $\dot{l}_{\rm p,grav}=\Gamma_{\rm grav}/m_{\rm p}$. 
Another component that drives the planet migrate comes from the accretion onto the planet. The force associated with accretion is
\begin{equation}
    \bm F_{\rm acc}=\int_{\delta r<r_{\rm a}}\bm v d\dot{m}_{\rm p},
\end{equation}
where $\bm v$ is the fluid velocity in the inertial frame. 
The accretion torque is $\bm \Gamma_{\rm acc}=\bm r_{\rm p}\times\bm F_{\rm acc}$, which changes the angular momentum by $m_{\rm p}\dot{l}_{\rm p,acc}\simeq\bm r_{\rm p}\times\int_{\rm \delta r<r_{\rm a}}(\bm v-\bm v_{\rm p})d\dot{m}_{\rm p}$. 
The total change rate of the specific angular momentum is thus $\dot{l}_{\rm p}=\dot{l}_{\rm p,grav}+\dot{l}_{\rm p,acc}$.
Generally, the evolution of $a_{\rm p}$ is dominated by the gravitational torque. 
We can calculate the cumulative gravitational torque 
\begin{equation}
    \Gamma_{\rm grav,<r}=\int_{r_{\rm in}}^r\int_\phi\int_\theta \rho\frac{\partial\Phi_{\rm p}}{\partial\phi} r^2 \sin\theta drd\phi d\theta.
\end{equation}
Note that the $\frac{\partial\Phi_{\rm p}}{\partial\theta}$ component is vanished considering of the 
vertical symmetry of the disk. 
The total gravitational torque on the planet is $\Gamma_{\rm grav}=\Gamma_{\rm grav,<r_{\rm max}}$
\par
The evolution of the semimajor axis for a planet is directly related to its specific angular momentum, which is 
\begin{equation}
    \frac{\dot{a}_{\rm p}}{a_{\rm p}} = 2\frac{\dot{l}_{\rm p}}{l_{\rm p}}-\frac{\dot{M}_*}{M_*},
\end{equation}
where $l_{\rm p}=\sqrt{GM_*a_{\rm p}}$ is the specific angular momentum of the circular planet. 
The second term associated with central stellar accretion is usually negligible. Here we have assumed that the planet is fixed on a circular orbit.


\section{results}\label{sec:result}

A summary of the model parameters and simulation results is shown in Table \ref{tab:para}. We explore the effect of planet mass for different viscosities and disk aspect ratios on the accretion and migration dynamics.

\subsection{Intermediate Planet Mass Case}

\begin{figure*}
    \centering
    \includegraphics[width=0.8\linewidth]{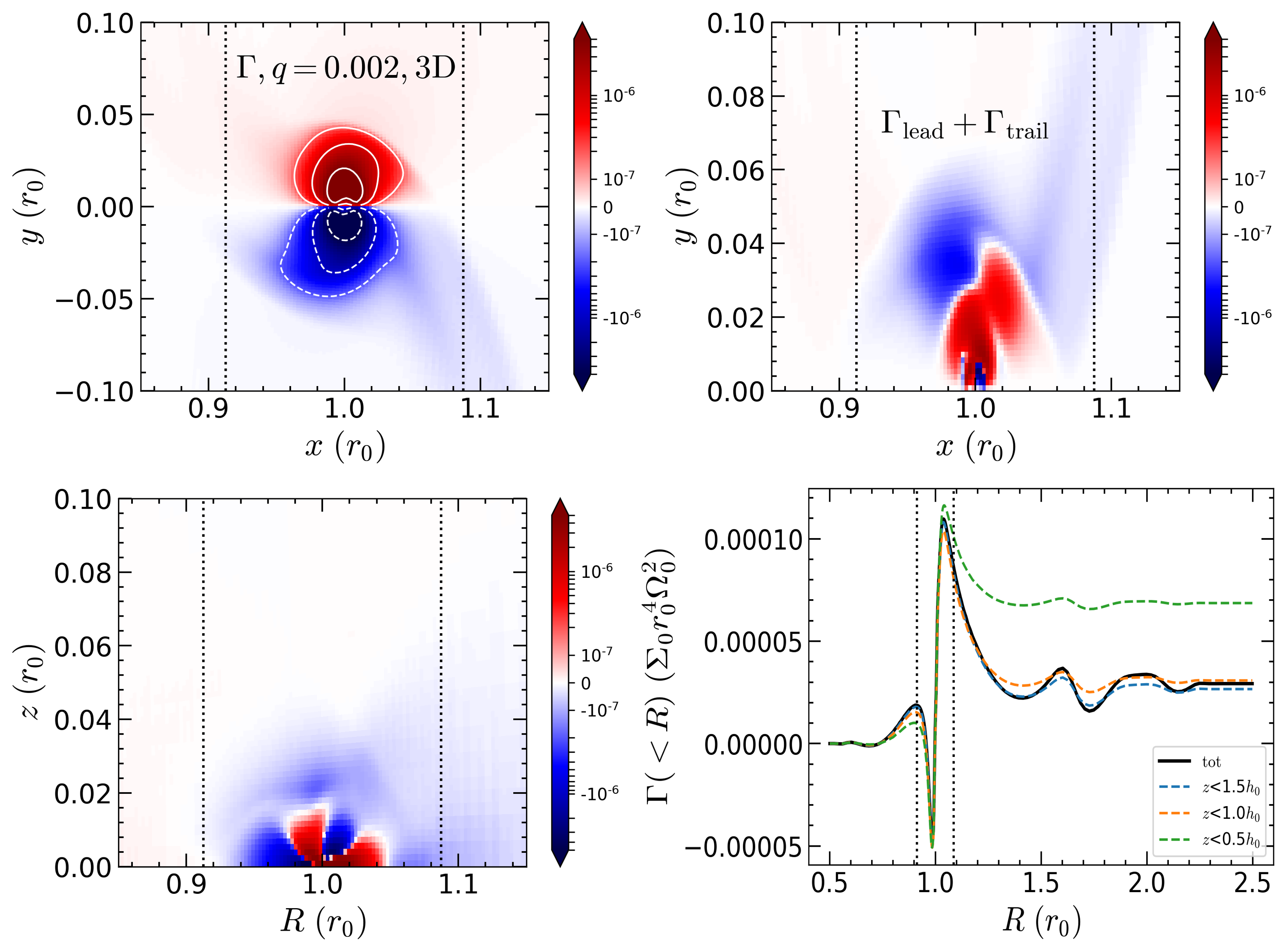}
    \caption{Spatial distribution of gravitational torque on planet for $q=2\times10^{-3}$(model \texttt{mp2e3}). 
    The upper left panel shows the torque around the planet in Cartesian Coordinate. 
    The red and blue part represent positive and negative torque, respectively. 
    In order to unveil the asymmetry between the leading horseshoe (upper) and trailing horseshoe (lower) region, we "fold" and sum them in the upper right panel, i.e., $\Gamma(r,\phi)+\Gamma(r,-\phi)$. 
    The lower left panel shows the azimuthally averaged torque distribution in the $r-z$ coordinate. 
    The lower right panel shows the cumulative radial torque distribution integrated within three different height $z$. 
    The total torque can be obtained from the value of $r=r_{\rm out}$. The vertical dashed line correspond to the region of Hill radius $R_{\rm H}$. 
    The planet is located in $x=r_0, y=0, R=r_0, z=0$. It can be seen that the positive torque mostly comes from the co-orbital region, and the negative torque from the outer spiral arm.
    }
    \label{fig:torque_mp2e3}
\end{figure*}

\begin{figure*}
    \centering
    \includegraphics[width=0.8\linewidth]{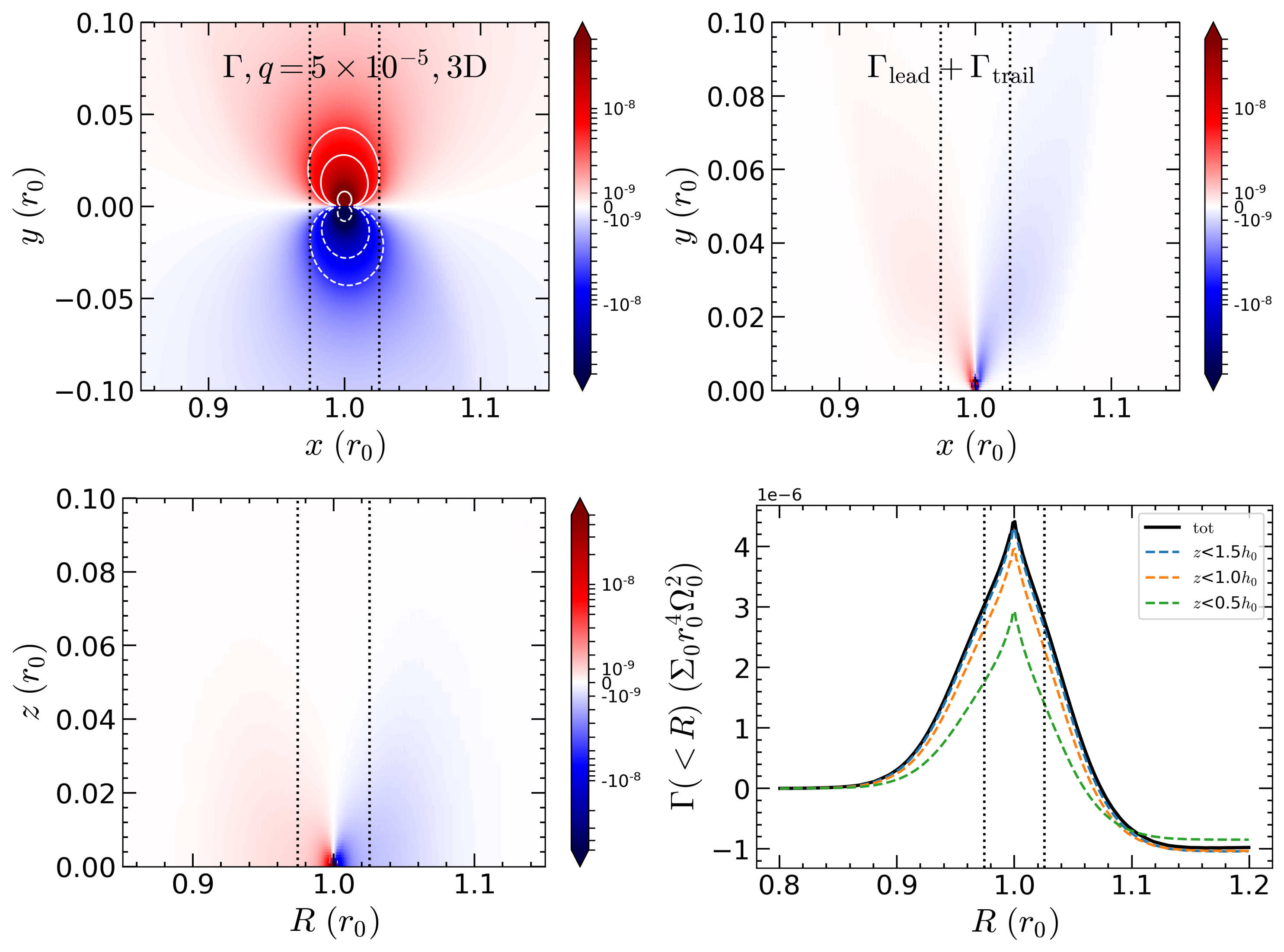}
    \caption{Same as Figure \ref{fig:torque_mp2e3}, but for the mass ratio $q=5\times10^{-5}$. We only shows the torque near the planet ($0.8<r<1.2$) in lower right panel since the torque contribution from the region far away from the planet location is negligible.}
    \label{fig:torque_mp5e5}
\end{figure*}

We first explore the case with $h_0=0.05$ and the mass ratio $q=2\times10^{-3}$. 
This case is labeled as model \texttt{mp2e3} in Table \ref{tab:para}. 
At the outer boundary of disk, $r_{\rm max}=2.5r_0$, the viscous timescale $t_{\rm vis}\simeq4000$ planetary orbits.  
In order to understand the role of planet accretion on the dynamical evolution of the planet, we evolve the global disk long enough to reach the quasi-steady state. 
The time evolution of the planet's accretion rate is shown in the upper row of Figure \ref{fig:1}. 
After approximately 1000 planetary orbits, $\dot{m}_{\rm p}$ stabilizes to a quasi-steady value of approximately $\sim8.9\times10^{-4}\Sigma_0r_0^2\Omega_0$, which is slightly below the disk supply rate $\dot{m}_{\rm d}$. It can be seen that $\dot{m}_{\rm p}$ barely evolves after 2000 orbits, indicating the achievement of a quasi-steady state.

\subsubsection{Dynamics of Planet}

The evolution of the semi-major axis (i.e., the migration rate) $\dot{a}_{\rm p}/a_p$ over time for the model \texttt{mp2e3} is shown in the middle and lower panels of Figure \ref{fig:1}. 
We show the migration rate due to the gravitational torque and the accretion torque on the planet separately. 
With accretion included, the planet migration rate $\dot{a}_{\rm p}/a_{\rm p}$ due to the gravitational torque becomes positive (blue line), and slightly negative due to the accretion torque (green line). 
The net migration rate, shown in the lower panel, is positive, indicating that the planet undergoes outward migration.
This is well consistent with our previous studies presented in \citet{Li2024}. 

Similarly to $\dot{m}_{\rm p}$, $\dot{a}_{\rm p}/a_{\rm p}$ reaches an asymptotic limit $\dot{a}_{\rm p}/a_{\rm p}\simeq32\ \dot{m}_{\rm d}/M_{*}$, suggesting a steady state of global disk. 
Since a sink hole is applied near the planet, some fraction of the gravitational torque from the sink cell may be overestimated, we therefore remove the torque within the sink cell, and compare the result (orange dashed line) with full gravitational torque in the middle panel. 
These two prescriptions show very similar profiles, suggesting that the torque from the sink hole is negligible.

\subsubsection{Torque on Planet}

The gravitational torque provides the main contribution to migration. 
Figure \ref{fig:torque_mp2e3} (upper left panel) displays the vertically integrated torque $\Gamma(r,\phi)$ after reaching quasi-steady state. 
The inner and outer spiral arms generate positive and negative Lindblad torque, respectively, result in a net negative torque. 
However, the total torque contributed by the CPD region dominates over that from spiral arms. 
The upper right panel of  Figure \ref{fig:torque_mp2e3} shows the vertically integrated, azimuthal differential torque around the planet defined as $\Gamma_{\rm asy}=\Gamma(r,\phi)+\Gamma(r,-\phi)$, such that only the asymmetry component of torque is left. 
Planet's accretion can destroy the symmetry of CPD, the positive torque from the leading side ($\phi \geq 0$) is stronger than the negative torque from the trailing side ($\phi \leq 0$). 
Consequently, this asymmetry drives the planet to migrate outward. 
The lower right panel of Figure \ref{fig:torque_mp2e3} shows the cumulative radial torque distribution, in which the torque undergoes drastic changes are between the dashed line (Hill radius), confirming the expectation that CPD contributes a decisive positive torque.


\subsection{Low and High Planet Mass Cases}

\begin{figure*}
    \centering
    \includegraphics[width=0.8\linewidth]{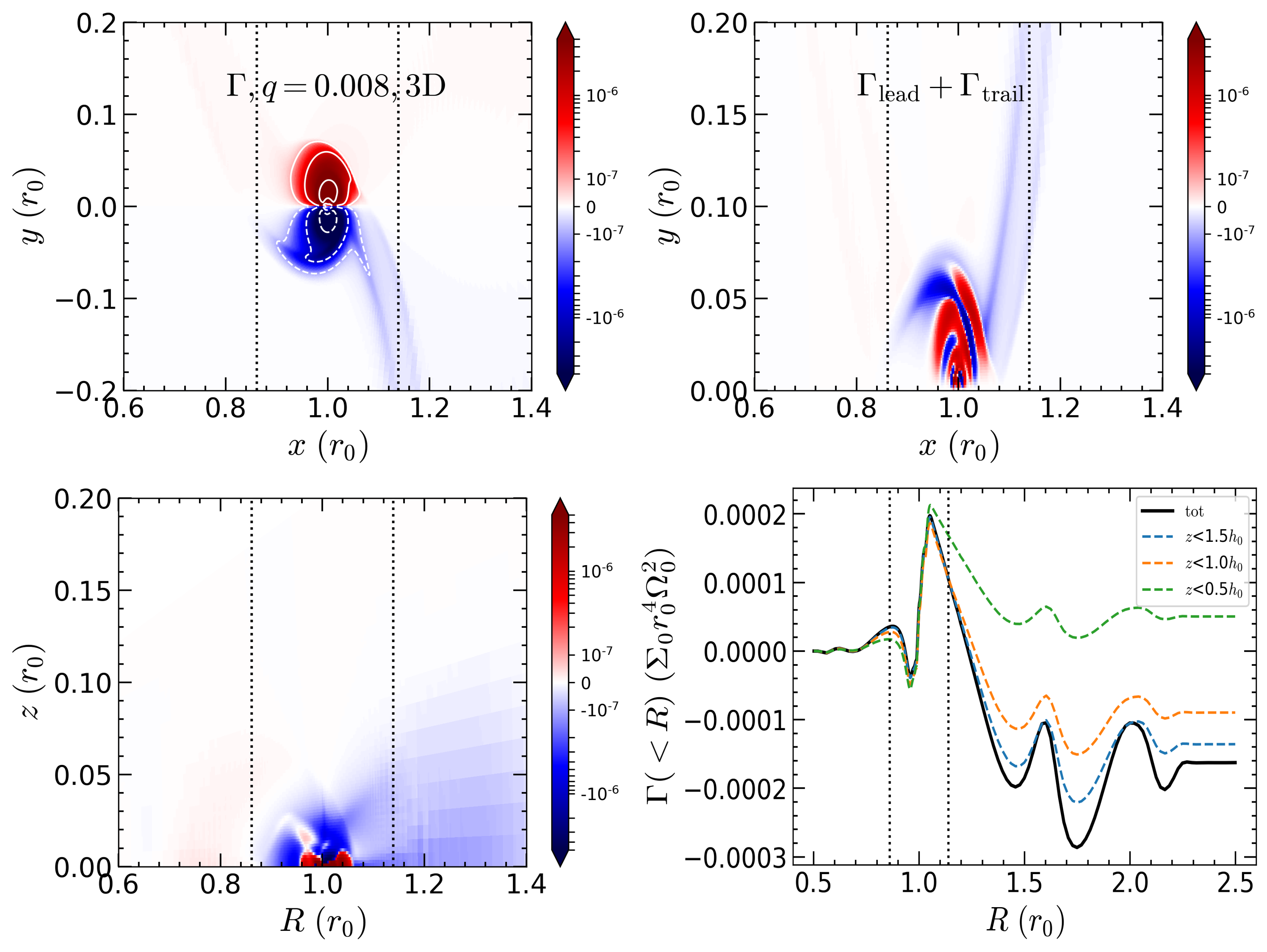}
    \caption{Same as Figure \ref{fig:torque_mp2e3}, but for the mass ratio $q=8\times10^{-3}$. The asymmetry between the leading and trailing sides around planet is not so pronounced, and the upper right panel shows an alternating pattern of positive and negative values. The figure shows a larger space to highlight the Lindblad torque that become significant at this mass ratio.
    }
    \label{fig:torque_mp8e3}
\end{figure*}

\begin{figure}
    \centering
    \includegraphics[width=\linewidth]{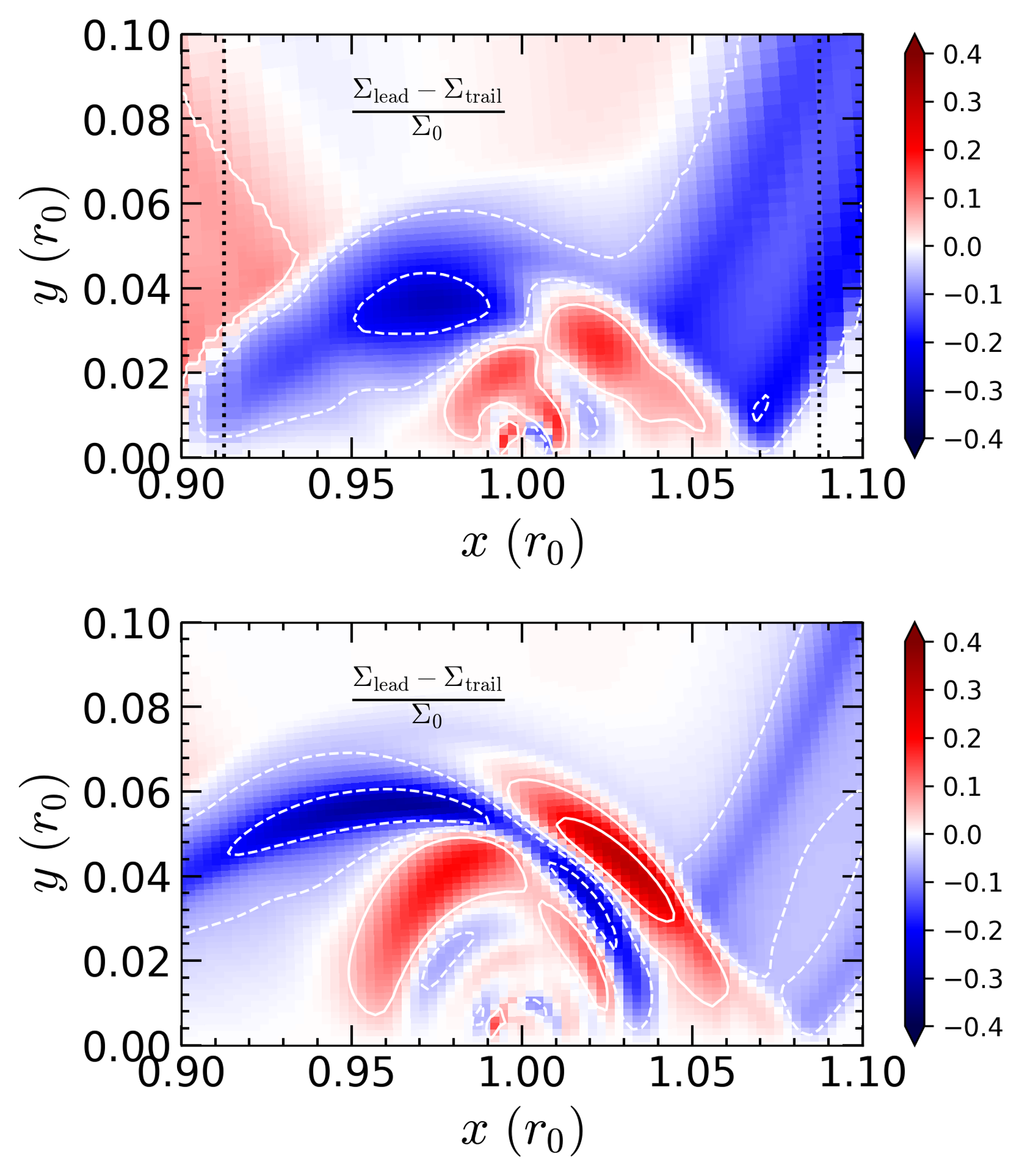}
    \caption{The distribution of normalized surface density for model \texttt{mp2e3} (upper panel) and \texttt{mp8e3} (lower panel), where we fold the distribution in the trailing side of planet to the leading side, i.e., $(\Sigma_{\rm lead}-\Sigma_{\rm trail})/\Sigma_0$. 
    The contour lines mark the values of 0.05 and 0.2, with dashed lines for negative value and solid lines for positive value.
    }
    \label{fig:sigma}
\end{figure}

As the planet mass decreases further, the gravitational capture of the accreted gas material becomes weaker which could result in a weaker accretion effect on the migration. 
To this end, we conduct a low mass ratio simulation with $q=5\times10^{-5}$. 
Such a small mass ratio planet has a Hill radius $R_{\rm H}=(q/3)^{\frac{1}{3}}r_{0}\sim0.026\ r_{0}$, and softening radius $\epsilon=0.1\ R_{\rm H}$. 
To resolve the reduced Hill sphere, we increased the static mesh refinement (SMR) level to 5 (instead of level 3 for intermediate mass ratio). 
Since the planet Hill radius $R_{\rm H}$ is now less than than the Bondi radius $R_{\rm B}$, i.e., $R_{\rm H}<R_{\rm B}$, this further necessitates the 3D simulations for the accreting planet.
We labeled the low mass ratio model as \texttt{mp5e5} in Table \ref{tab:para}.

The time evolution of accretion rate $\dot{m}_{\rm p}$ and migration rate $\dot{a}_{\rm p}/a_{\rm p}$ of model \texttt{mp5e5} is shown in the middle column of Figure \ref{fig:1}. 
It takes about 4000 orbits to reach quasi-steady state. 
As a result, the accretion rate $\dot{m}_{\rm p}$ is well below $\dot{m}_{\rm d}$. 
A significant amount of gas continues to flow past the planet into the inner disk without being accreted. 
The measured $\dot{a}_{\rm p}/a_{\rm p}$ due to the gravitational torque becomes negative, while the $\dot{a}_{\rm p}/a_{\rm p}$ due to accretion is slightly positive. 
The net $\dot{a}_{\rm p}/a_{\rm p}$ turns out to be negative, tend to recover to the classical type I migration rate.


For extremely small planet with $q_{\rm th}\lesssim\sqrt{3}/3$, where $q_{\rm th}\equiv q/h_0^3$ is the planetary thermal mass, its Hill sphere entirely embedded in the disk. Consequently, the surrounding gas accumulates into an envelope rather than a CPD, eliminating the asymmetry between the leading and trailing sides of the planet, as illustrated in Figure \ref{fig:torque_mp5e5}.
Meanwhile, the asymmetry between the inner spiral and outer spiral arms still persists, resulting in a slight negative torque. The gravitational torque is thus dominated by the Lindblad resonant instead of the asymmetric component of the CPD region.
Among all our models, the gravitational torque is consistently one to two orders of magnitude larger than the accretion torque, therefore, it is the negative Lindblad torque that drives the planet migrate inward.

At higher planetary masses, the gap deepens sufficiently to quench gas accretion \citep[e.g.,][]{Bryden2000,Li2021,Li2023}, potentially altering CPD dynamics and planet migration.
To extend to a higher mass ratio, we performed a simulation with $q=8\times10^{-3}$, labeled in Table \ref{tab:para} as model \texttt{mp8e3}. 
The right column of Figure \ref{fig:1} shows the accretion rate $\dot{m}_{\rm p}$ and migration rate $\dot{a}_{\rm p}/a_{\rm p}$ evolution over time. 
After the mass is increased to $q=8\times10^{-3}$, it takes about 1000 orbits to reach a quasi-steady state. 
The accretion rate $\dot{m}_{\rm p}$ of the high-mass planet is lower than that of the intermediate-mass planet, which is consistent with the expectation of the significant gap opening effect by giant planets.


As Figure \ref{fig:torque_mp8e3} shows, the azimuthal asymmetry of the horseshoe region still exists. We quantify this density asymmetry ($\varsigma\equiv (\Sigma_{\rm lead}-\Sigma_{\rm trail})/\Sigma_{0}$) more elaboratively around the planet for model \texttt{mp2e3} and model \texttt{mp8e3} in Figure \ref{fig:sigma}.  A clear asymmetry of the spiral arms around the CPD region can be identified. A positive (negative) $\varsigma$  corresponds to a positive (negative) gravitational torque. This asymmetry can thus be responsible for generating the net gravitational torque acting on the planet. Such an asymmetry is on the similar order of magnitude for model \texttt{mp2e3} and model \texttt{mp8e3}.
However, the positive gravitational torque from such kind of asymmetric structures cannot balance the negative differential Lindblad torque due to the $q^{2}$ dependence of differential Lindblad torque.

We note that when the planet is more massive (e.g., $q\gtrsim1\times10^{-2}$), the Lindblad regions will behave quite differently, e.g., an eccentric gap would be formed which could suppress the outer Lindblad torque. 
This is associated with an accretion outburst, disk and planet eccentricity excitation.
Under this condition, outward migration could be expected for very massive non-accreting planet cases \citep{Dempsey2021}.

However, in our 3D cases, the outer disk has yet to be excited to be eccentric.
As a result, the outer Lindblad torque is still strong enough to drive the planet to migrate inward for the most massive model \texttt{mp8e3} with $q_{\rm th}=64$.
This reason could be two-fold. One is that the inclusion of the planet accretion further deplete the inner Lindblad region such the positive Lindblad torque from the inner Lindblad resonance is suppressed.  Another reason could be related to the fact that the transition to an eccentric disk in 3D simulations requires an even higher planet mass compared to 2D counterparts \citep{Li2023}.

\subsection{Modifications of Type I and II Migration with Concurrent Accretion}

\begin{figure*}
    \centering
    \includegraphics[width=0.6\linewidth]{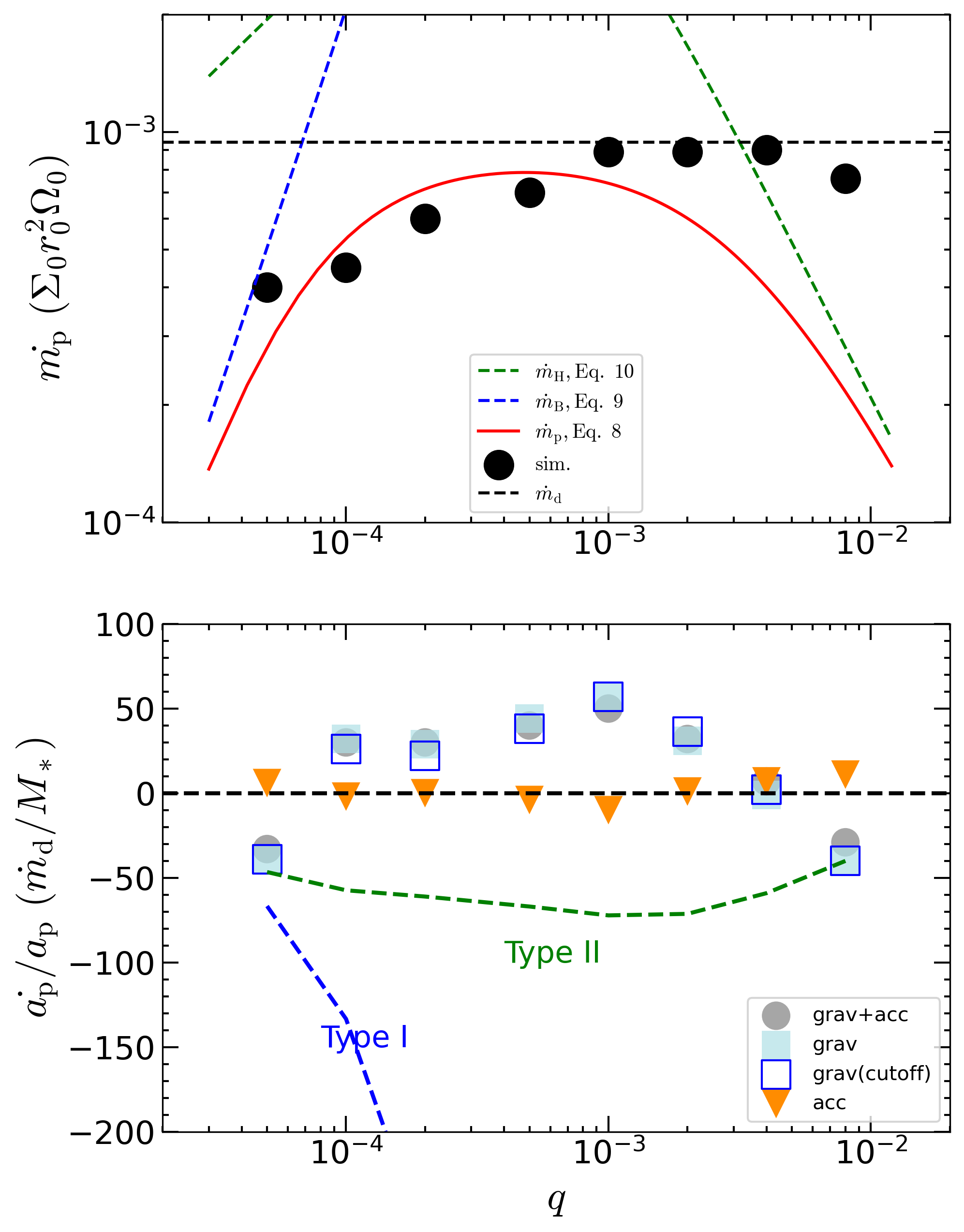}
    \caption{The planetary accretion rate $\dot{m}_{\rm p}$ and migration rate $\dot{a}_{\rm p}/a_{\rm p}$ as a function of mass ratio $q$ in the upper and lower panel, respectively.
    In the upper panel, the dashed line shows the disk accretion rate from outer boundary $\dot{m}_{\rm d}$. 
    The red, blue, green lines shows the theoretical accretion rates based on Equation (\ref{eq:mpdot}), (\ref{eq:Bondi}), and (\ref{eq:Hill}), respectively. 
    In the lower panel, the blue and green dashed lines show the type I and type II migration rates where the global disk depletion effect due to gap opening has been considered. 
    All points in this figure are from 3D simulation, with fixed viscosity $\alpha=0.04$, and scale height $h_0=0.05$.
    }
    \label{fig:adot_mdot}
\end{figure*}

In order to study the relationship between mass ratio and migration, we carry out a few additional runs with different values of mass ratio. 
The results are summarized in Table \ref{tab:para} and Figure \ref{fig:adot_mdot}. 
We show the relationship between $q$ and $\dot{m}_{\rm p}$ in the upper panel of Figure \ref{fig:adot_mdot}. 
The viscosity is maintained at $\alpha=0.04$, and the disk aspect ratio is $h_0=0.05$.

To understand the accretion onto the planet, we adopt the theoretical accretion rate of the planet as \citep{Li2023}:
\begin{equation}
    \dot{m}_{\rm p}=\frac{\dot{m}_{\rm p,loc} \dot{m}_{\rm d}}{\dot{m}_{\rm p,loc}+\dot{m}_{\rm d}},
    \label{eq:mpdot}
\end{equation}
where $\dot{m}_{\rm p,loc}$ is the unimpeded gas accretion rate onto the planet with infinite disk gas supply.
\par
For the less massive planet with $q_{\rm th}\lesssim\sqrt3/3$, the accretion follows  the formalism of Bondi accretion \citep{Li2022,Choksi2023,Li2023}:
\begin{equation}
    \dot{m}_{\rm p,loc}=\dot{m}_{\rm B}=\pi q_{\rm th}^2 h_{\rm 0}^3 \rho_{\rm p} r_{\rm 0}^3 \Omega_{\rm 0},
    \label{eq:Bondi}
\end{equation}
where $\rho_{\rm p}$ is the characteristic gas density at the planet location. 
While the massive planet with $q_{\rm th}\gtrsim3$ exhibits Hill accretion formalism \citep{Li2023,Li2024}:
\begin{equation}
    \dot{m}_{\rm p,loc}=\dot{m}_{\rm H}=k\left(\frac{q_{\rm th}}{3}\right)^{2/3}h_{\rm 0}^3\rho_{\rm p} r_{\rm 0}^3 \Omega_{\rm 0},
    \label{eq:Hill}
\end{equation}
\citet{Choksi2023} find a pre-factor of $k\simeq 9$ while \citet{Li2023} show that $k\simeq \pi$.  
Here we adopt this pre-factor of $k=3$, and will further discuss this issue later.

\par
For the intermediate mass regime between Bondi accretion and Hill accretion, \citet{Ida2025} found a formula that combines Equation (\ref{eq:Bondi}) and (\ref{eq:Hill}), and describes the accretion rate of the planet as (see also \citealt{TanigawaTanaka2016}),
\begin{equation}
    \frac{\dot{m}_{\rm p,loc}}{\dot{m}_{\rm d}}=\xi=\frac{1}{3\pi \alpha \left(1+K^\prime\right)}\left(\sqrt{2/\pi}q_{\rm th}^{-2}+\frac{1}{k}\left(\frac{q_{\rm th}}{3}\right)^{-2/3}\right)^{-1}, 
    \label{eq:zeta}
\end{equation}
where $K^\prime=0.04 q_{\rm th}^2 h_{0} \alpha^{-1}$.
The parameter $\xi$ can be regarded as the strength of planetary accretion. 
As $\xi$ increases, the planet captures a larger fraction of gas supplied from the outer boundary. 
Those not captured by the planet will pass through the co-orbital region and eventually accreted by the host star. 

In the upper panel of Figure~\ref{fig:adot_mdot}, we show the theoretical planetary accretion rate as a function of the mass ratio for the case with $h_0=0.05$ and $\alpha=0.04$. 
We can see that the accretion rate for the mass ratio below $q\simeq10^{-3}$ can be satisfactorily in agreement with the simulation results. However, for the more massive planets with a pre-factor of $k=3$, especially for $q=0.008$, the predicated accretion is a factor of a few smaller than those of simulation results. We have confirmed that simulated accretion rates cannot decline significantly with an extending simulation time to match the theoretical predication. One may suspect that the disk eccentricity could be excited for a massive planet, which may enhance the accretion onto the planet \citep{Kley2006,Li2021}.
However, we have not observed strong accretion outbursts, which suggests a lacking of a significant disk eccentricity excitation \citep{Li2021,Li2023}, as further confirmed by the disk gap profiles.  One possible reason for this discrepancy is that the accretion onto the more massive planet will behave more similarly to the circumbinary accretion rather than Hill accretion predicated here. \citet{Duffell2020b} found that the the secondary accrete at a rate $\sim10$ times of the primary for a mass ratio $q=0.01$, which suggests that the secondary would accrete almost all the supply from the circumbinary disk \citep[see review by][]{Lai2023}. We would thus expect that the accretion onto the more massive planet will transit to the binary case when $q$ is large enough, even though the exact boundary is still not clear \citep[but see][]{DOrazio2016}.

In all the high-resolution simulations of accreting planets (including those presented here), 
their orbits are fixed for computational convenience.  This assumption introduces a potential 
paradox for planets with $q_{\rm th} \gg 1$.   
In reality, planets undergo migration \citep{Lin1996} which could naturally resolve the inconsistency between the expected suppression of ${\dot m}_{\rm p}$ in a severely depleted gap and the unimpeded ${\dot m}_{\rm p} \simeq {\dot m}_{\rm d}$ in the simulated high-mass model
(top panel, Fig.~\ref{fig:adot_mdot}). 
We discuss this effect in detail below.

Combining the continuity and momentum equations, the azimuthal equation of motion for the steady state is \citep{Lin1986,Li2024,Laune2024}
\begin{equation}
-rv_{r}\Sigma \frac{\partial v_{\phi}r}{\partial r} + \frac{\partial}{\partial r}\left(r^{3}\nu\Sigma  \frac{\partial \Omega}{\partial r}\right) -r\Sigma \frac{\partial\Phi_{\rm p}}{\partial \phi} =0. 
\end{equation}

Integrating the above equation, we find
\begin{equation}
-r^{2}v_{r}\Sigma v_{\phi} + r^{3}\nu\Sigma  \frac{\partial \Omega}{\partial r} + \int_{r}^{r_{\rm max}}\left(r\Sigma \frac{\partial\Phi_{\rm p}}{\partial \phi}\right){\rm d}r \equiv \dot{J}.
\label{eq:Jr}
\end{equation}
Here we have assumed an axisymmetric disk and neglected the $v_{r}$ term of the viscous stress. The first term in the l.h.s. of the above equation is the advected angular momentum flux, the second term is the viscous flux, and the third term is related to the gravitational torque from the planet on the disk.  $\dot{J}$ is the total inward angular momentum flux. Note that $2\pi \int_{r_{\rm in}}^{r_{\rm out}}\left(r\Sigma \frac{\partial\Phi_{\rm p}}{\partial \phi}\right){\rm d}r=\Gamma$ is the total torque acting on the planet from the disk. If we neglect the accretion torque on the planet $\dot{a}_{\rm p}$, then we have $\dot{a}_{\rm p}/a_{\rm p}=2\Gamma/m_{\rm p}a_{\rm p}^{2}\Omega_{\rm p}$.  
Considering the outer disk ($r>r_{\rm p}$), the last term in the l.h.s of Eq.~\ref{eq:Jr} is the outer Lindblad torque when the CPD torque becomes less important for an inward migrating giant planet. This will provide the dominant negative torque for the inward migration.


In the steady state solution with our torque free boundary condition, the angular momentum flux for the inner disk ($r<r_{\rm p}$) is imposed to be $\dot{J}_{\rm in}=0$ \citep{Li2024}. The angular momentum flux transferred to the accreting planet is then solely determined by the angular momentum flux of the outer disk $\dot{J}_{\rm out}$.
We note that the magnitude of the third term in the l.h.s of Eq.~\ref{eq:Jr} is usually much less than the first and second terms since $|m_{\rm p}\dot{a}_{\rm p}/a_{\rm p}| \ll \dot{m}_{\rm d}$ for the moderate gap cases, as shown in Table~\ref{tab:para}. 
However, the inward angular momentum flux for the outer disk $\dot{J}_{\rm out}$ can be suppressed when the negative gravitational torque from a massive giant planet increases to compensate for the advected angular momentum flux (i.e., $|m_{\rm p}\dot{a}_{\rm p}/a_{\rm p}| \sim \dot{m}_{\rm d}$). In such a case with a smaller $\dot{J}_{\rm out}$, we could thus expect a smaller $\dot{m}_{\rm p}$ for a migrating planet.
Simulations for these released planets will be explored in details in the near future.

\begin{figure*}
    \centering
    \includegraphics[width=\linewidth]{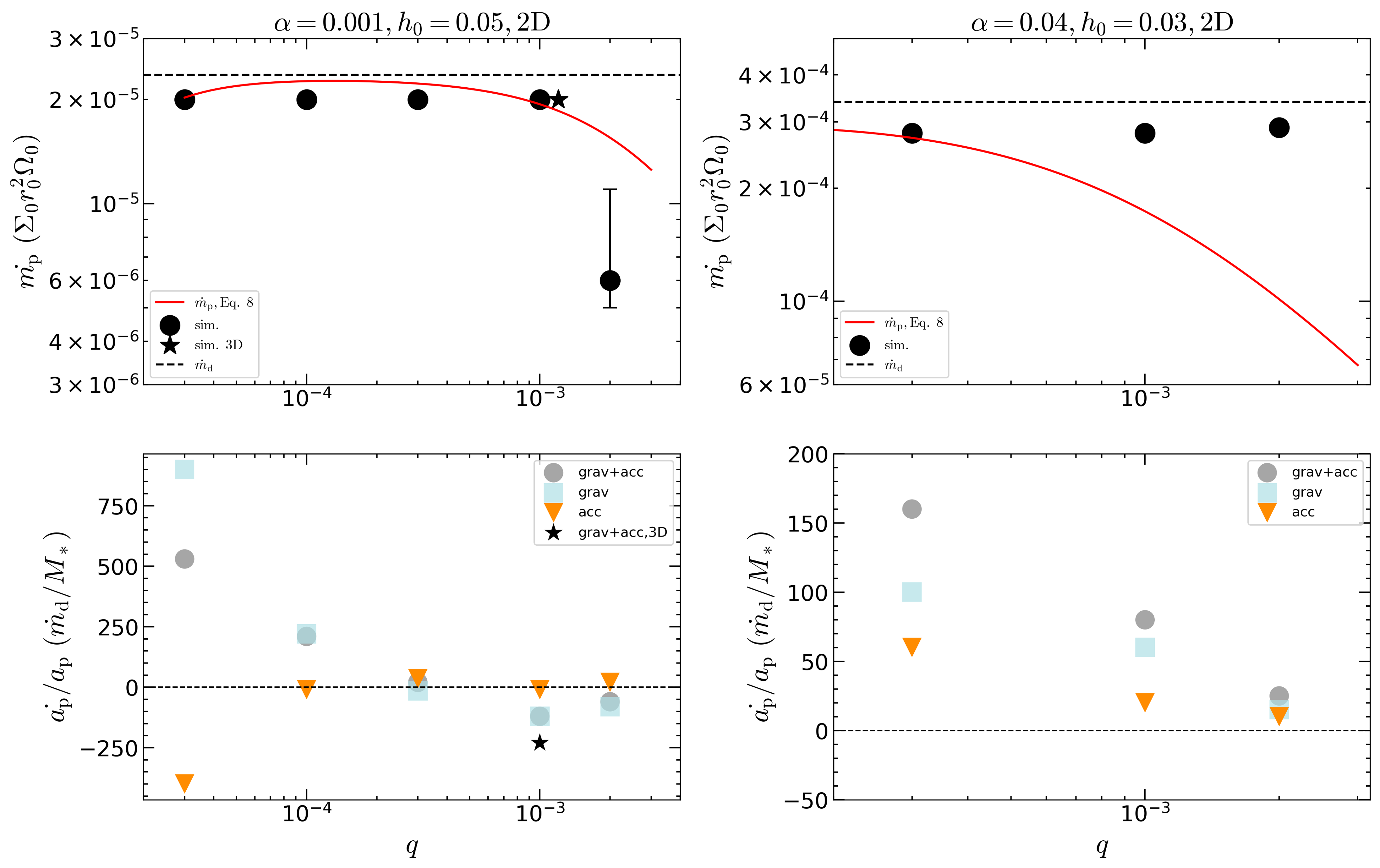}
    \caption{Same as Figure \ref{fig:adot_mdot}, except for the left column with $\alpha=0.001$ and for right column with $h_0=0.03$. All points shown in this figure are from 2D simulations, except for the pentagram representing a 3D simulation. The accretion rate for model \texttt{mp2e3a} varies with time with the error bar indicating its variability amplitude.
    }
    \label{fig:2d}
\end{figure*}

The lower panel of Figure \ref{fig:adot_mdot} shows the migration rate $\dot{a}_{\rm p}/a_{\rm p}$ as a function of the mass ratio $q$. 
Migration is dominated by gravitational torque instead of the accretion component. 
It is insensitive to whether we have neglected the sink cell region. 
At this value of viscosity $\alpha=0.04$ and scale height $h_0=0.05$, planets with mass ratios in the range $1\times10^{-4}\lesssim q \lesssim 4\times10^{-3}$ exhibit outward migration. 
Planetary accretion significantly alters the structure of the disk, thereby influencing planetary migration. 
As a result, the migration of accreting intermediate-mass planets deviate substantially from the theoretical predictions of Type I or Type II regimes. 
The low-mass planet, such as the model \texttt{mp5e5}, barely perturbs the global disk and only captures a smaller fraction of the disk accretion rate, its migration recovers to Type I. 
Migration rate of Type I is \citep{Tanaka2002}:
\begin{equation}
    \frac{\dot{a}_{\rm I}}{a}\sim  \frac{q_{\rm th}}{3\alpha h_{0}} \frac{\dot{m}_{\rm d}}{M_{\rm *}}.
    \label{eq:type I}
\end{equation}
High-mass ratio planet perturbs the disk so significant that a gap will be opened near the planet's orbit. 
The gap decreases the disk's density and planet's migration rate by a factor of $1/(1+K^\prime)$\citep{Kanagawaetal2015}. 
In addition, accretion of the planet causes the depletion of global disk density, which can also decrease the migration rate by a factor of $f_{\rm pass}=1/(1+\xi)$ \citep{TanigawaTanaka2016}. 
Finally, the type II migration rate, such as model \texttt{mp8e3}, is:
\begin{equation}
    \frac{\dot{a}_{\rm II}}{a}\sim \frac{f_{\rm pass}}{(1+K^\prime)} \frac{\dot{a}_{\rm I}}{a}.
    \label{eq:type II}
\end{equation}

The predictions of the type I and modified type II migration rates are shown in Figure~\ref{fig:adot_mdot} to compare with our simulation results. In the intermediate-mass ratio range, where the accretion effect is strongest, there exists an additional positive torque contribution from the CPD region which is not included in the formula above. A detailed analytic formula of the migration rate for the accreting planet is provided in \citet{Ida2025}.

We find that the positive torque in the intermediate mass ratio regime (or the accreting-dominated regime) shows a similar magnitude as the inward migration torques (the regime dominated by the differential Lindblad torque). This could be due to that the azimuthal asymmetry in the CPD region shows a magnitude on the order of the disk aspect ratio $h_0$ as shown in Figure~\ref{fig:sigma}, which is roughly consistent with the predicated azimuthal asymmetry $\sim h_0$ shown in \citet{Ida2025}. 
As the radial asymmetry for the Lindblad torque shows a similar scaling with $h_0$, this could explain the similar magnitude of positive torque in the accreting-dominated region.

\subsection{Simulations With Different Viscosities and Disk Aspect Ratios}

We have also performed additional 2D simulations to further explore the parameter dependence of the planetary accretion and migration as shown in Figure \ref{fig:2d}.
Here we have considered the cases with $\alpha=10^{-3}$, $h_{0}=0.05$ (left panels) and $\alpha=0.04$, $h_{0}=0.03$ (right panels) for different mass ratios $q$.

The accretion rates are generally consistent with the theoretical predication although there are some discrepancies for models with a larger mass ratio.
For migration dynamics, we find that a Jupiter-mass planet will migrate inward for a viscosity of $\alpha=10^{-3}$ in our previous work.
Here, we find that outward migration can also occur under the condition of $q\lesssim3\times10^{-4}$ as shown in the left panel of Figure \ref{fig:2d}.
We have performed a 3D simulation with $\alpha=10^{-3}$ and $q=10^{-3}$, which confirms that our 2D simulations are consistent with 3D results.
Planets with large masses gradually approach the Type II (inward) migration rate, while those with masses below this threshold migrate outward. 
We expect that planets with smaller masses will revert to Type I migration, which has not been simulated in this work.
However, simulating such small-scale planets requires 3D simulations with higher resolutions and longer simulation times, which are extremely challenging. Furthermore, such a low mass ratio with $q\lesssim3\times10^{-5}$ may have not experienced runaway gas accretion yet, but it could be applied to sBH accretion embedded in AGN disks.

As shown in the right panel of Figure \ref{fig:2d}, we further reduced the disk scale height to $h_0=0.03$ but still with a disk viscosity parameter of $\alpha=0.04$. 
The planet generally migrates outward for our explored planet mass ranges of $3\times10^{4}\lesssim q\lesssim2\times10^{-3}$.  There is tendency that the more massive planet will migrate inward.

We integrate the results of different values of $\alpha$ and $h$, transforming the relationship between migration rate and mass ratio into a relationship with $K^{\prime}$. This allows us to determine the range of $K^{\prime}$ that causes the planetary outward migration, which is $K^\prime_{\rm acc} \lesssim K^\prime \lesssim K^\prime_{\rm gap}$, where $K^\prime_{\rm acc}\simeq 0.03(h_{0}/0.05)$ and $K^\prime_{\rm gap}\simeq 50$. The derivation of this condition is discussed in detail in a companion paper by \citet{Ida2025}.

\section{Summary and Discussions}\label{sec:summary}

We followed \cite{Li2024} and performed high-resolution 3D/2D hydrodynamical simulations to study the migration of giant accreting planets in disks over a broader parameter space. 
We flexibly adjusted the SMR level and mass-growing time in different models to meet the numerical resolution and stability requirements for both low-mass and high-mass planet's simulations. 
By extending the mass ratio to a broader range, we identify the upper and lower mass limits that allow for planetary outward migration.

We explore the dependence of planetary migration on planet-to-star mass ratios under various viscosity and disk scale heights. 
We conclude that planetary accretion leads to an outward migration trend, which is consistent with the findings of \cite{Li2024}. 
The accretion rates of low-mass planets are naturally small and belong to Type I (inward) migration. High-mass planets can open a gap on the disk, which hinders their own accretion, aligning with Type II (inward) migration.

In contrast, intermediate-mass planets have the highest accretion rates capped by the disk supply, and such a substantial accretion effect can lead to outward migration. 
This outward migration should therefore occur within specific mass ratio bounds dictated by disk parameters, namely $\alpha$ and $h_{0}$.
For example, when $\alpha=0.04$ and $h_0=0.05$, the mass ratio range for outward-migrating planets is measured to be $1\times10^{-4}\lesssim q\lesssim 4\times10^{-3}$. 
When the viscosity is adjusted to $\alpha=0.001$, the upper limit is measured as $q\lesssim 3\times10^{-4}$. 
When the disk scale height is adjusted to $h_0=0.03$, the upper limit is measured as $q\simeq 2\times10^{-3}$.
But no lower limit is found in both cases, as it becomes challenging to simulate such small planets.

We have already explored a vast parameter space and obtained a wealth of simulation results. 
Finding a quantitative formula to explain and predict the planetary migration rates under different conditions has become a feasible task. 
Based on the simulation data above, we have identified the conditions that allow for planetary outward migration, $K^{\prime}_{\rm acc}\lesssim K^\prime\lesssim K^\prime_{\rm gap}$, where $K^\prime_{\rm acc}\simeq0.03(h_{0}/0.05)$ and $K^\prime_{\rm gap}\simeq50$ \citep{Ida2025}. 
Within this regime, the accretion of the planet significantly deviates from the classical theoretical framework for migration. 
Specifically, when $K^\prime$ is less than $K^\prime_{\rm acc}$, the planet's perturbation on the disk--and consequently its accretion rate--is negligible, leading to Type I (inward) migration. 
Conversely, when a deep gap is opened in the disk under the condition of $K^\prime>K^\prime_{\rm gap}$, leading to suppressing on planetary accretion, this situation corresponds to Type II (inward) migration. 
Thus, the termination of planetary mass growth is synonymous with the transition from
outward to inward migration.
With such a new concurrent migration and accretion pattern, we expect that the architecture of planets be potentially modified, which could be reconciled with the formation of different populations of exoplanets, especially cold Jupiters.
The details of the analytical formula for the outward migration regime is provided in a separate work \citep{Ida2025}.

Several aspects of this work can be improved in future studies. First, for the parameter survey explored so far, we focus on the planet in a circular orbit. Observationally, exoplanets can be in eccentric orbits, especially for super-Jupiter cases. 
How the eccentricity of an accreting planet excites the eccentricity of the disk and subsequently affects the migration of the planet is an important factor to consider in the next step. 
Second, among all of our simulations, the planet's orbital semi-major axis is fixed to ensure the local steady state around the planet to be established quickly. In reality, the planet will migrate in the disk, and planetary lively migration can alter measurements such as accretion and migration rate. Whether this effect significantly affects our results remains to be determined in future work. 
Third, with the new accretion and migration formula, the dynamics of multiplanet system could also be modified. Specifically, one related question is how this could affect the mean-motion resonance dynamics of the planet pair embedded in PPDs.

Additionally, our model can be regarded as a special case of a binary system with an extreme mass ratio. When the mass ratio is close to $q\sim1$, the binary orbit is expected to expand but could shrink in a colder disk \citep{Munoz2019, Munoz2020, Moody2019, Duffell2020b, DOrazio2021,Lai2023}.
The accretion onto the secondary for the circumbinary disk is also different from the planetary accretion \citep{Duffell2020b,Lai2023}. 
\citet{DOrazio2016} observed a transition from an annular ring to a central cavity at $q\simeq0.04$ based on 2D hydrodynamical simulations, although their analysis omitted torque and orbital evolution effects.
Identifying the boundary between binary and planet accretion, and migration could be a potential direction for future study.

\acknowledgements
We would like to thank the anonymous referee for helpful comments and suggestions. J.P. and Y.P.L are supported in part by the Natural Science Foundation of China (grants 12373070, and 12192223), the Natural Science Foundation of Shanghai (grant NO. 23ZR1473700). S.I. is supported by JSPS Kakenhi grant 21H04512 and 23H00143. The calculations have made use of the High Performance Computing Resource in the Core Facility for Advanced Research Computing at Shanghai Astronomical Observatory. 
Softwares: \texttt{Athena++} \citep{Stone2020}, \texttt{Numpy} \citep{vanderWalt2011}, \texttt{Scipy} \citep{Virtanen2020}, \texttt{Matplotlib} \citep{Hunter2007}.

\par

\bibliography{references.bib}{}
\bibliographystyle{aasjournalnolink}

\end{document}